\documentclass[12pt]{article}
\usepackage{graphicx}
\usepackage{amsfonts}

\begin{document}
\begin{titlepage}

\hfill{Preprint {\bf SB/F/03-313}} \hrule \vskip 2.5cm

\centerline{\bf Husimi's $Q(\alpha)$ function and quantum}
\centerline{\bf interference in phase space} \vskip 1.5cm
\centerline{D.F. Mundarain$^1$ and J. Stephany$^{1,2}$} \vskip 4mm

\centerline{$^1${\it  Departamento de F\'{\i}sica, Universidad
Sim\'on Bol\'{\i}var,}}
\centerline{\it Apartado 89000, Caracas
1080A, Venezuela.}

\centerline{$^2${\it  Departament of Physics and Centre for
Scientific Computing,} } \centerline{\it The University of
Warwick, Coventry, CV4 7AL, UK.}

\begin{abstract}
We discuss a phase space description of the photon number
distribution of non classical states which is based on Husimi's
$Q(\alpha)$ function and does not rely in the WKB approximation.
We illustrate this approach using the examples of displaced number
states and two photon coherent states and show it to provide an
efficient method for computing and interpreting the photon number
distribution . This result is interesting in particular for the
two photon coherent states which, for high squeezing, have the
probabilities of even and odd photon numbers oscillating
independently.
\end{abstract}
\vskip 2cm
\hrule
\bigskip
\centerline{\bf UNIVERSIDAD SIM{\'O}N BOL{\'I}VAR} \vfill
\end{titlepage}

\section{Introduction}
The oscillations in the photon number distribution are an
interesting feature common to various kind of light states, which
may be taken as a signal of  non classical behavior of those
states. They were first computed for squeezed states
\cite{Wheeler,Schleich} and interpreted as an interference effect
in phase space \cite{Wheeler,Schleich,Walls,Dowling}. They also
have been the subject of experimental
investigations\cite{Schiller} in connection with the properties of
Wigner's distribution. Since in the actual experimental situation
many factors like the detector properties may modify the
conclusions of the theoretical analysis it is of interest  to
understand better the physics involved and to develop our
intuition of how the oscillations may appear and eventually
disappear.

For our analysis we consider a single mode of the electromagnetic
field.  For this case, the picture of quantum interference of
states in phase space was developed as a generalization of the
Bohr-Sommerfeld description of quantum states as finite areas in
phase space \cite{Wheeler,Schleich,Walls,Dowling,Kim}. The total
area of each state is fixed by the normalization requirement
$\langle \psi|\psi \rangle =1$. The area associated with a Fock or
number state $|m>$ is a circular ring centered at the origin with
interior  radius $r_m^- = \sqrt{2 m}$ and exterior radius $r_m^+ =
\sqrt{2 m +2}$. The areas associated to coherent states and
squeezed states are obtained by displacing, or squeezing and
displacing the one associated to the vacuum state. The inner
product $<\psi|\phi>$ between two quantum states is then related
with the overlapping area of the states in the phase space. The
intersecting region may have in some interesting cases more than
one component. Since the probability amplitude $<m|\psi>$ is a
complex number and the overlapping areas are real, it results
natural to associate to each of these components a phase in order
to reproduce the quantum mechanical results. The probability
amplitude acquires then the following structure,
\begin{equation}\label{ec1a}
<\chi|\psi> = \sum_i \sqrt{A_{\chi\psi}^i(\psi)} e^{\phi_{\chi\psi}^i(\psi)}
\end{equation}
where  $A_{\chi\psi}^i(\psi)$ and $\phi_{\chi\psi}^i(\psi)$ are
the $i$-th component of the overlapping area and their assigned
phase respectively.

The presence of the areas   $A_{\chi\psi}^i(\psi)$ in this
expression is very natural and physically appealing although the
fact that what appears is the square root of the areas does not
allow any direct geometrical method to derive Eq.(\ref{ec1a}). On
the other hand the values that have to be chosen for the phases
$\phi_{\chi\psi}^i(\psi)$ are not evident from the geometry of the
pase space. Dowling ({\it et al.})\cite{Dowling} work out a quite
general equivalent of Eq. (\ref{ec1a}) for the probability
amplitude of the eigenstates of two different hamiltonian
operators using the WKB approximation. They  obtained an explicit
expression of the phases where it is possible to recognize a
geometrical content. Oscillations in the photon statistics for
displaced states, two photon squeezed states and for squeezed
number states may also be studied with this methodology  but this
approach is limited by the validity of the WKB approximation
\cite{Kim,Mundarain}.

Some time after, Milburn \cite{Milburn} show that the interference
effects in phase space may also be understood  by considering the
properties of Husimi's function
\begin{displaymath}
Q(\alpha)=\frac{1}{\pi}|<\alpha|\psi>|^2 \ \ ,
\end{displaymath}
for the state under study. This author
notes that the over-completeness of the coherent states allows to
rewrite the probability amplitude $\langle m|\psi\rangle $ as a
phase space integral,
\begin{equation}
\label{ec1}
<m|\psi> = \frac{1}{\pi} \int d^2 \alpha <m|\alpha><\alpha|\psi>.
\end{equation}
The functions  $<m|\alpha>$ and $<\alpha|\psi>$ could be
interpreted as the phase space probability amplitudes for the
states   $|m>$ and $|\psi>$ respectively. Each of these functions
is proportional to the function $Q(\alpha)$ of the corresponding
state. Then,
\begin{equation}
<m|\alpha> = \sqrt{\pi Q_m(\alpha)} e^{i \phi_m(\alpha)}
\end{equation}
\begin{equation}
<\alpha|\psi> = \sqrt{\pi Q_{\psi}(\alpha)} e^{i \phi_{\psi}(\alpha)}
\end{equation}
One may then approximate the integral in (\ref{ec1}), by
restricting the domain of integration to the phase space region
where the product of the two probability amplitudes is appreciably
different of zero and by identifying the regions were the $Q$
function concentrates with the phase space regions assigned to the
states, this approximation is equivalent to consider the
integration domain as the overlapping areas between the different
states. Following this line of thought in this paper we discuss in
detail the photon number distributions for the displaced number
states and the two photon coherent states.  We show that this
approach allows to identify the areas and phases for the phase
space description without rendering in the WKB approximation. This
results is interesting in particular for the two photon coherent
states  which has, for high squeezing, the probabilities of even
and odd photon numbers oscillating independently.

\section{Photon statistics for displaced number states}

Let us consider first the example of the displaced number states.
They are defined through the action of the
displacement operator  $D(\beta)= \exp(\beta a^{\dagger}-\beta^*a)$
on the  Fock states $|n\rangle$,
\begin{equation}
|n,\beta \rangle   = D(\beta) |n\rangle\ \ .
\end{equation}
For simplicity we work with real $\beta$.

We compute the photon distribution using the method of generating functions.
Consider a coherent state $|\alpha\rangle=D(\alpha)|0\rangle$
with real $\alpha$. We have,
\begin{equation}
D(\beta)|\alpha\rangle = |\alpha+\beta\rangle = e^{-|\alpha|^2/2}
\sum_{n=0}^{\infty} \frac{\alpha^n}{\sqrt{n!}} D(\beta) |n\rangle
\end{equation}
Then,
\begin{equation}
\langle m|D(\beta)|n \rangle = \frac{1}{\sqrt{n!}}  \left\{\left(
\frac{\partial}{\partial \alpha}\right)^n \left( e^{|\alpha|^2/2}
\langle m|\alpha+\beta\rangle\right) \right\}_{\alpha=0}
\end{equation}
with
\begin{equation}
\langle m|\alpha+\beta \rangle  =\frac{1}{\sqrt{m!}}
e^{-\frac{|\alpha+\beta|^2}{2}} (\alpha+\beta)^m .
\end{equation}
As in other cases the photon number distribution $P_{mn}(\beta) =
|\langle m|n,\beta\rangle |^2$ is oscillating. (see Figure
(\ref{fig6}) below)

To develop the phase space description consider the phase space amplitudes,
\begin{eqnarray}
\langle m|\alpha \rangle & =& \frac{1}{\sqrt{m!}} e^{-|\alpha|/2} \alpha^m\nonumber\\
& = &|\langle m|\alpha \rangle |\quad e^{i \phi_m(\alpha)}
\label{malfa}
\end{eqnarray}
and
\begin{eqnarray}
\langle \alpha|n,\beta \rangle & =& \frac{1}{\sqrt{n!}} e^{\frac{\beta \alpha^*-\alpha \beta^*}{2}} e^{-\frac{|\alpha-\beta|^2}{2}} (\alpha^*-\beta^*)^n \nonumber\\
&= &|\langle \alpha|n,\beta \rangle |\quad e^{- i
\phi_{n,\beta}(\alpha)} .
\label{alfanbeta}
\end{eqnarray}
The phases here are better expressed in terms of the real an
imaginary parts of $\alpha=x+iy$ and take the form
\begin{equation}
\phi_m(\alpha) =  m \arctan \left(\frac{y}{x}\right)
\label{fim}
\end{equation}
and
\begin{equation}
\phi_{n,\beta}(\alpha)  = \frac{\alpha \beta^* -\beta
\alpha^*}{2 i}+ n  \arctan \left( \frac{y}{x-\beta}\right)\ \ .
\label{finbeta}
\end{equation}

\begin{figure}[th]
\includegraphics[scale=0.5,angle=-90]{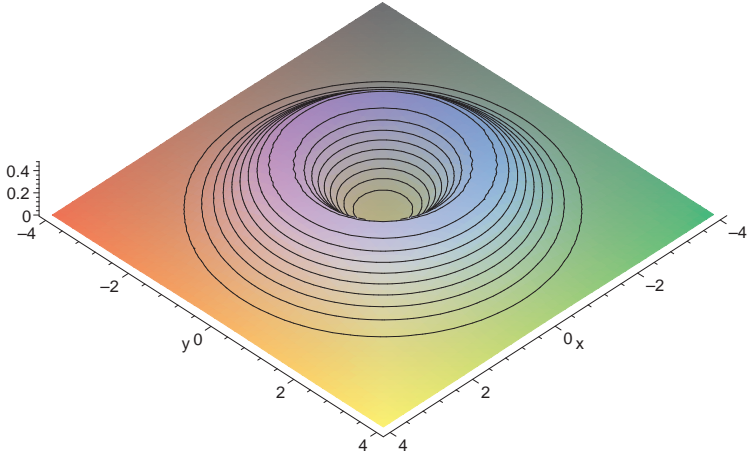}
\caption{$|<m|\alpha>|$ for a Fock state with $m=3$ } \label{fig1}
\end{figure}

In Figure(\ref{fig1}) we show the phase space probability
amplitude module \mbox{$|\langle m|\alpha \rangle|$} for a number
state with $m=3$. This function reminds the  Bohr-Sommerfeld  ring
associated with the corresponding number state except for the fact
that here the mean radius is $|\alpha | = \sqrt{m}$. This suggests
to associate the number state $|m\rangle$ the circular ring
centered at origin and located between the radii $\sqrt{m-1/2}$
and $\sqrt{m+1/2}$. Note also that the phase space amplitude
$|\langle \alpha | n,\beta \rangle| $ for the displaced number
state $|n,\beta \rangle$ is obtained by displacing the amplitude
$|\langle \alpha | n\rangle| $.

\begin{figure}[hbt]
\includegraphics[scale=0.4,angle=-90]{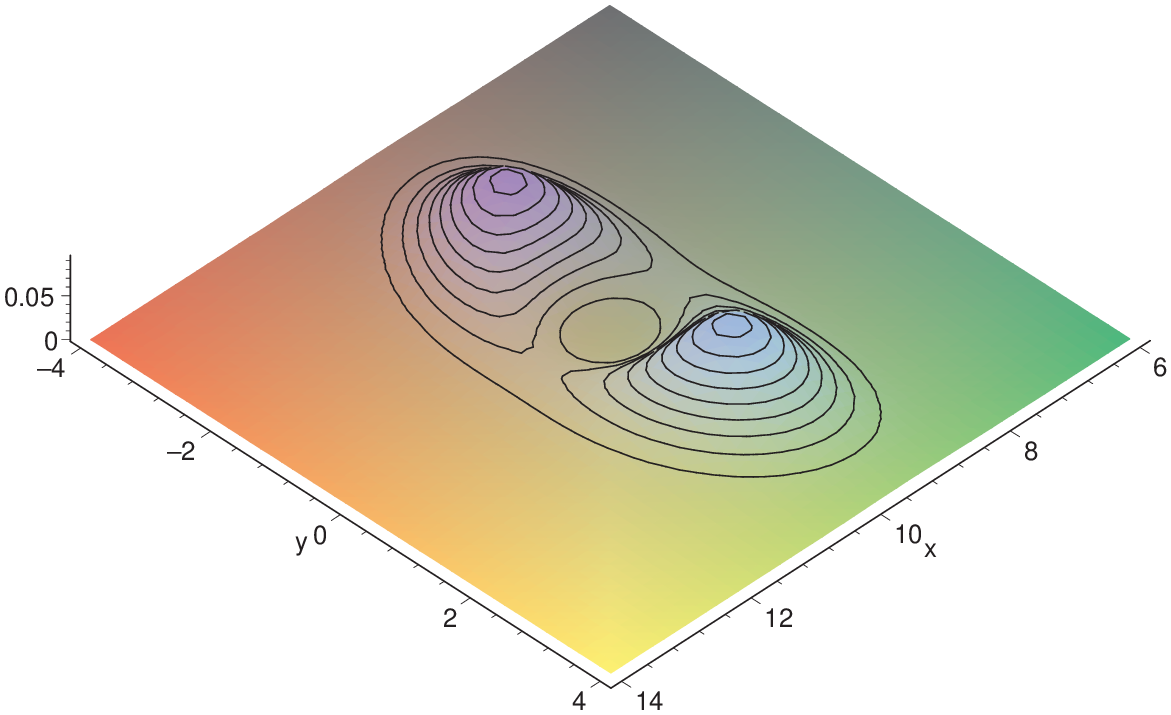}
\caption{ \mbox{$|<m|\alpha><\alpha|n,\beta>|$}  for a Fock state
with $m=100$ and a displaced state with $n=3$ and $\beta=10.1$ }
\label{fig2}
\end{figure}

\begin{figure}[hbt]
\centerline{\includegraphics[scale=0.4]{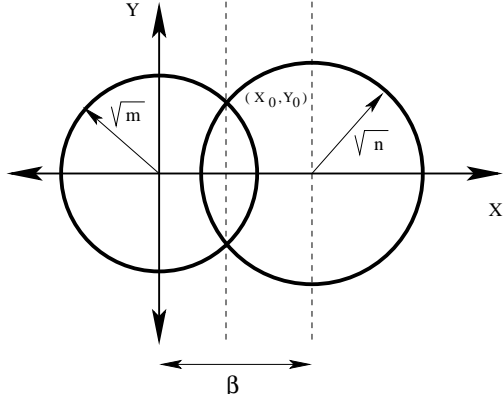}}
\caption{Intersection points in phase space} \label{fig3}
\end{figure}

In Figure (\ref{fig2}) we show the product
$|<m|\alpha><\alpha|n,\beta>|$ for $m=100$, $n=3$ and $\beta=
10.1$. Observe that  there are two overlapping regions where both
terms are appreciably not vanishing. But note also that the
intersection areas with this prescription are not in the same
position were they would appear if working with the
Bohr-Sommerfeld bands. Now they may be localized in the
intersection points of a circumference of radius $\sqrt{m}$
centered at origin and a second circumference of radius $\sqrt{n}$
displaced by $\beta$, as shown in Figure(\ref{fig3}). The
intersection points are then given by $\alpha_+= x_0+iy_0$ and $
\alpha_- = x_0-iy_0$, with
\begin{equation}
x_0= \frac{\beta^2+m-n}{2 \beta}
\label{ekis0}
\end{equation}
and
\begin{equation}
y_0= \sqrt{m-x_0^2}
\label{lle0}
\end{equation}

\begin{figure}[h]
\centerline{\includegraphics[scale=0.4]{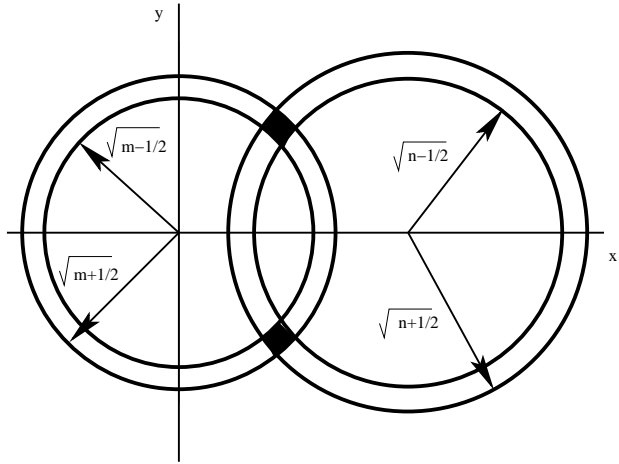}}
\caption{Overlapping areas between the Fock state $|m>$ and the
displaced number state $|n,\beta>$} \label{fig4}
\end{figure}

Now we approximate the phase space integral of Equation (\ref{ec1}) for the
photon number probability amplitude  $\langle m|n,\beta\rangle$ by taking
instead of (\ref{malfa}) and (\ref{alfanbeta}) uniform contributions on the intersection
of the rings described above with the angles (\ref{fim}) and (\ref{finbeta}) evaluated at
the points given by  (\ref{ekis0}) and (\ref{lle0}). The probability amplitude
has then the structure,
\begin{equation}
<m|n,\beta> = \frac{\sqrt{{\it A}_{mn}}}{\pi} e^{i
\psi}+\frac{\sqrt{{\it A}_{mn}}}{\pi} e^{-i \psi_{n,\beta}^{(m)}}
\end{equation}
where
\begin{eqnarray}
\psi_{n,\beta}^{(m)}& = & \phi_m(\alpha_+) - \phi_{n,\beta} (\alpha_+) \nonumber\\
&=&m \arctan \left(\frac{y_0}{x_0}\right) -n \arctan \left(
\frac{y_0}{x_0-\beta}\right) -\beta y_0
\label{angle1}
\end{eqnarray}
and  ${\rm A_{mn}}$ are the overlapping areas between the rings
corresponding to the number and displaced number states as shown
in  Fig.(\ref{fig4}).

\begin{figure}[h]
\centerline{\includegraphics[scale=0.4]{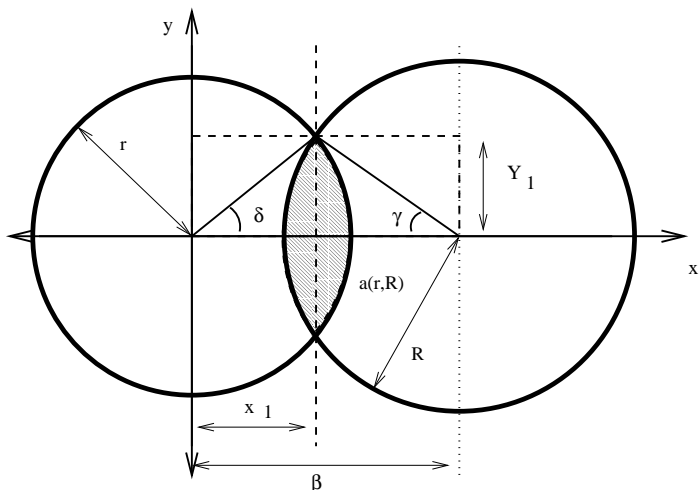}}
\caption{Internal areas used to define $A_{mn}$} \label{fig5}
\end{figure}

The computation of this areas is analogous to the one presented in
\cite{Walls,Dowling} for the Bohr-Sommerfeld strips and is given
by,
\begin{eqnarray}
A_{mn}&=& \frac{1}{2} \left( a(\sqrt{m+1/2},\sqrt{n+1/2})- a(\sqrt{m-1/2},\sqrt{n+1/2})\right.\nonumber\\
&&\left.- a(\sqrt{m+1/2},\sqrt{n-1/2})+ a(\sqrt{m-1/2},\sqrt{n-1/2})\right)
\end{eqnarray}
where $a(r,R)= r^2 \delta +R^2 \gamma - \beta Y_1$ is the internal
area between the two circular paths with radii $r$ and $R$ as
shown in Figure (\ref{fig5}), with $(X_1=
\frac{r^2+\beta^2-R^2}{2\beta},Y_1=\sqrt{r^2-X_1^2})$ the
intersection point of the circumferences, $\delta =
\arctan\left\{\frac{Y_1}{X_1}\right\}$ and $\gamma =
\arctan\left\{\frac{Y_1}{\beta-X_1}\right\}$.

Figure (\ref{fig6}) show the exact and approximate results of
$P_{mn}$ for a displaced state with $n=3$ and $\beta =10.1$
showing a good qualitative and quantitative behavior for the
values for which the assigned rings actually overlap.

\begin{figure}[ht]
\includegraphics[scale=0.4,angle=-90]{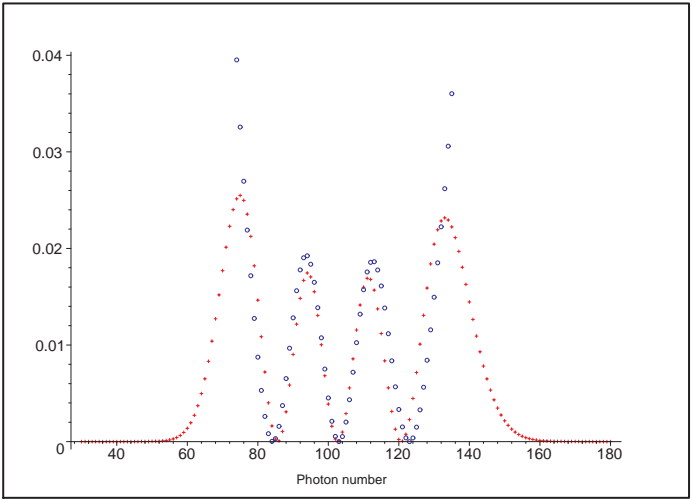}
\caption{Photon statistics $P_{mn}$ for a displaced number state
$|n,\beta>$ with $n=3$ and $\beta=10.1$. Exact computation:
crosses. Approximation:circles. } \label{fig6}
\end{figure}

\begin{figure}[ht]
\includegraphics[scale=0.4,angle=-90]{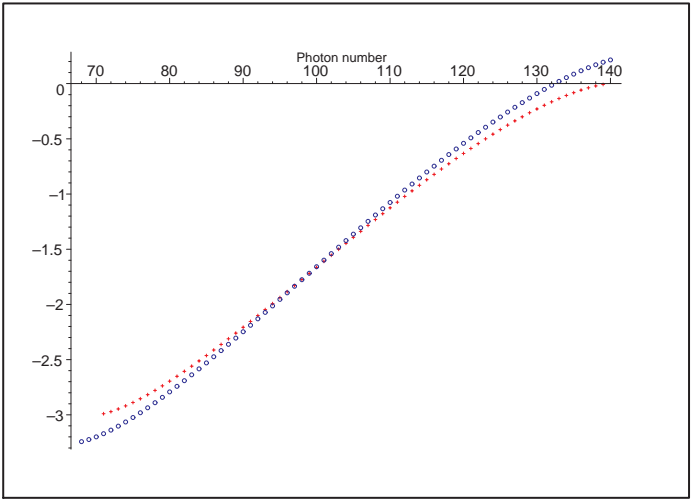}
\caption{Comparison between the phases. Crosses: $(\psi+n
\pi)/\pi)$. Circles: $(\psi^{WKB}+n \pi)/\pi$.} \label{fig7}
\end{figure}

The comparison  of the result in  Eq. (\ref{angle1})  which has
the direct geometrical interpretation
\begin{equation}
\psi_{n,\beta}^{(m)} = a(\sqrt{m},\sqrt{n})-n \pi
\end{equation}
with the one obtained by Dowling {\it et al} using the WKB
approach is shown in Figure (\ref{fig7}).

The WKB phase is given by \cite{Dowling,Merzbacher}
\begin{eqnarray}
\psi^{WKB}&=& -\left( m+1/2\right) \arcsin \left[ \frac{x_c(m,n)}{\sqrt{2m+1}}\right]
 +\left( n+1/2\right) \arcsin \left[ \frac{x_c(m,n)-\sqrt{2}\beta}{\sqrt{2n+1}}\right]\nonumber\\
&-&\frac{\sqrt{2}\beta}{2}\sqrt{2m+1-x_c(m,n)^2}-\frac{(n-m)\pi}{2} +\frac{\pi}{4}
\end{eqnarray}
where
\begin{equation}
x_c(m,n) = \frac{m-n}{\sqrt{2}\beta}+\frac{\sqrt{2}\beta}{2}
\end{equation}
The phases differ sensibly for zero and high photon number where
the WKB approximation fails to reproduce the distribution
\cite{Mundarain}.

\section{Two photon coherent states}

As a second illustration let us consider the
Two photon coherent states. These are obtained by the application of the
squeezing operator  $S(r) = \exp\left\{ \frac{r}{2} a^2 -
\frac{r}{2} (a^{\dagger})^2\right\}$ on the coherent state $|\alpha\rangle$.
That is,
\begin{equation}
|\beta,r\rangle = S(r) |\beta \rangle
\end{equation}
Again we take  $\beta$ and $\xi$ real.
The photon statistics is given in this case by \cite{Yuen,Albano},
\begin{eqnarray}
P_n = |\langle n|\beta,r\rangle|^2& =& \frac{(\tanh (r))^n}{2^n n! \cosh (r)} \exp \left\{ \beta^2 (\tanh(r)-1) \right\}\nonumber\\
&&\times \left|H_n \left( \frac{\beta}{\sqrt{2 \cosh(r) \sinh (r)}}\right)\right|^2
\end{eqnarray}
with   $H_n(x)$ the order $n$ Hermite polynomial.

\begin{figure}[ht]
\includegraphics[scale=0.5,angle=-90]{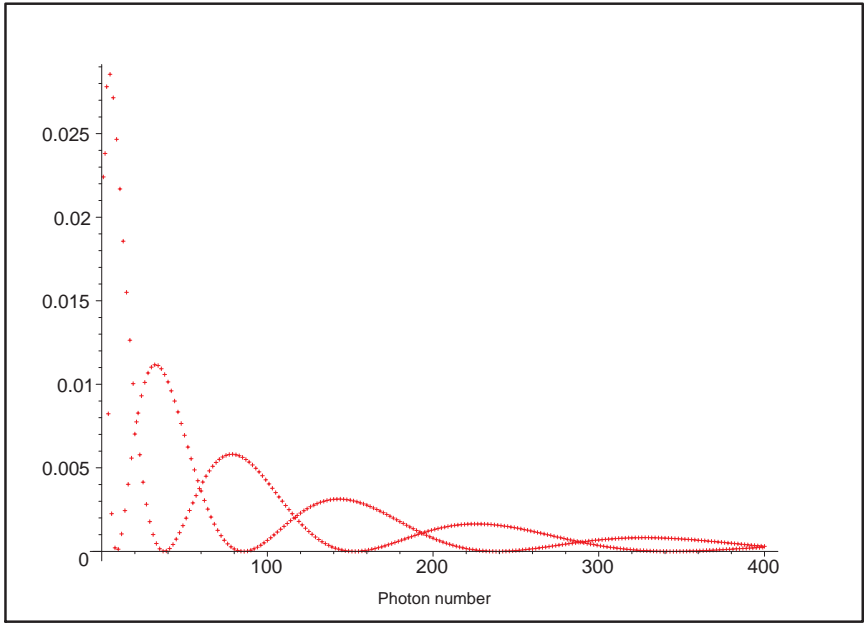}
\caption{Photon statistics for a two photon coherent state with
$\beta = 5.1$ and $r=3$.} \label{fig8}
\end{figure}

For  $r<<1$ the photon statistics resembles that of a coherent
state (although in fact the statistics is sub-poissonian). As the
squeezing increases, they appear the oscillations in the
distribution characteristic of the phase space interference. But
then for higher values of   $r$ ( see Figure (\ref{fig8}) ) the
distribution develops different oscillating behaviors for odd and
even photon numbers an effect which, as we show below, may be
understood in terms of phase space interference. Following the
same line as in the last section take Equations (\ref{malfa}) and
(\ref{fim}) and consider,
\begin{eqnarray}
\langle \alpha|\beta,r \rangle & =&\sqrt{{\rm sech} (r)}\exp \left\{-\frac{1}{2}(|\alpha|^2+\beta^2)+\alpha^* \beta {\rm sech} (r)\right.\nonumber\\
&&\left.-\frac{|}{2}((\alpha^*)^2-\beta^2)\tanh (r) \right\} \nonumber\\
& = &|\langle \alpha|\beta,r \rangle |\quad e^{i \phi_{\beta,r}(\alpha)}
\end{eqnarray}
which define
\begin{equation}
  \phi_{\beta,r}(\alpha) = - y \beta  {\rm sech}(r) + x y \tanh (r)\ \ .
\end{equation}

\begin{figure}[ht]
\includegraphics[scale=0.4]{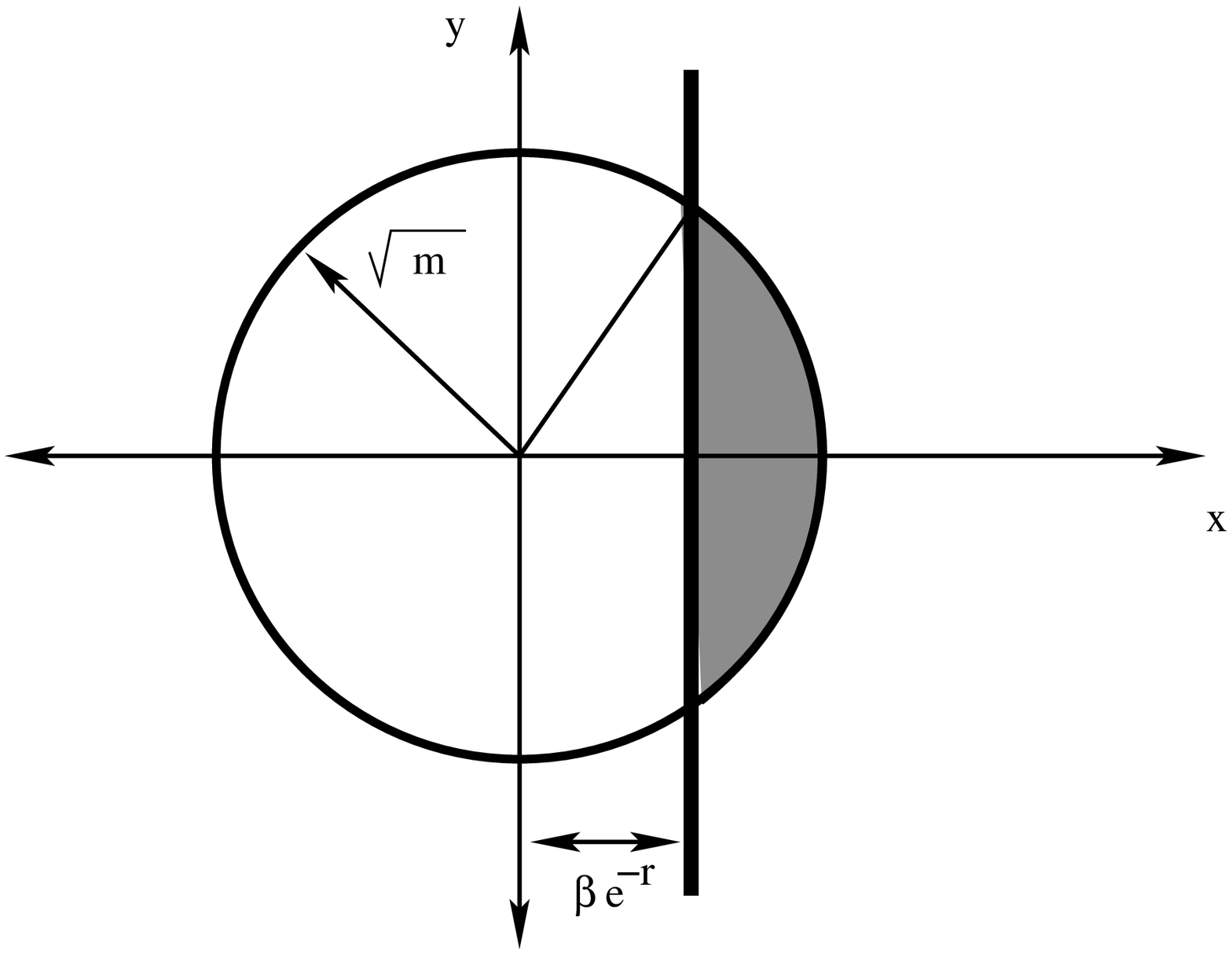}
\caption{Intersecting points which the phase for $<m|\beta,r>$}
\label{fig9}
\end{figure}
\begin{figure}[ht]
{\includegraphics[scale=0.4]{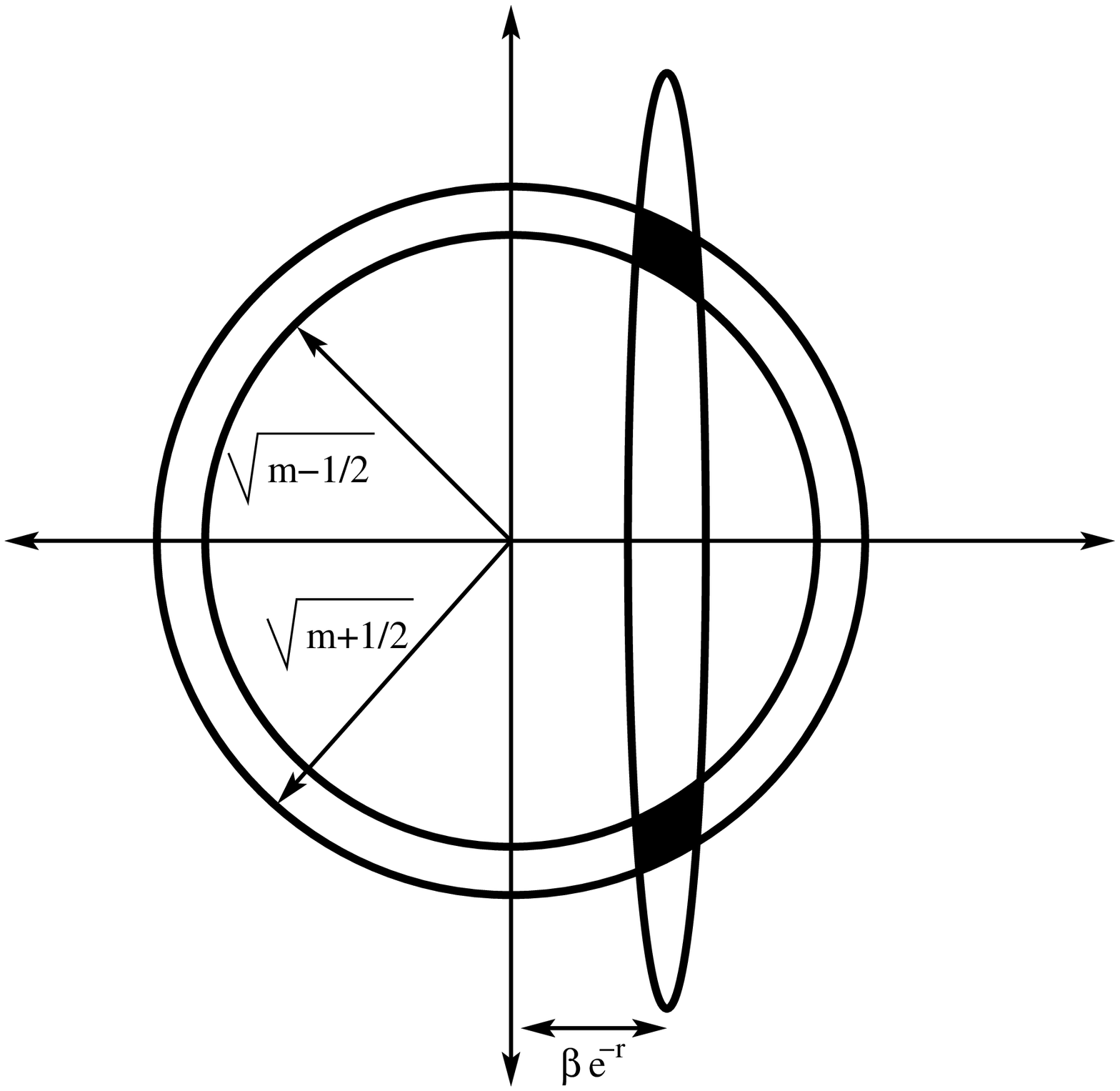} \caption{Overlapping areas
for $<m|\beta,r>$} \label{fig10}}
\end{figure}

It is not difficult to show that the phase space probability amplitude
$|\langle \alpha|\beta,r \rangle |$ of the two photon coherent
state concentrate on an ellipse  centered in  $\beta \exp\{-r\}$,
with one semi-axis of magnitude  $e^{-r}$ oriented through the $X$ axis  and the
other semi-axis of magnitude $e^r$ oriented through the $Y$ axis. We represent the state
by the internal region of the ellipse,
\begin{equation}
(x-\beta e^{-r})^2 e^{2r}+y^2 e^{-2r} = 1 .
\end{equation}
of area  $\pi$. For high values of  $r$ the ellipse approximates
to a vertical line.  In this limit the intersection points with
the circumference of radius $\sqrt{m}$ (which we are using to
represent the number states) are $\alpha_+ = X_2+iY_2$ and
$\alpha_- = X_2-iY_2$, with
\begin{eqnarray}
X_2 =\beta \exp\{-2r\}\nonumber\\
Y_2 = \sqrt{m-X_2^2} \ .
\label{ekis2}
\end{eqnarray}

As in the previous case we may write the probability amplitude
$\langle m|\beta,r \rangle $ as:
\begin{equation}
\langle m|\beta,r\rangle  =
\frac{\sqrt{A_{m}}}{\pi}e^{i\psi_{\beta,r}}+
\frac{\sqrt{A_{m}}}{\pi}e^{-i\psi_{\beta,r}^{(m)}}
\end{equation}
where the phases evaluated at the intersection points are given by
\begin{eqnarray}
\psi_{\beta,r}^{(m)} &=& \phi_m(\alpha_+)+\phi_{\beta,r}(\alpha_+)\nonumber\\
&=& m \arctan \left\{\frac{Y_2}{X_2} \right\} -Y_2 \beta {\rm sech}(r) +X_2 Y_2 \tanh (r)\label{ekis2bc}
\end{eqnarray}
and $A_m$ is one the shadowed areas in  Figure (\ref{fig10}). It
has the value
\begin{equation}
A_m = \Delta y \Delta x
\end{equation}
with
\begin{displaymath}
\Delta y = \sqrt{m+1/2-\beta^2 e^{-2r}} -\sqrt{m-1/2-\beta^2 e^{-2r}}
\end{displaymath}
\begin{displaymath}
\Delta x = 2 e^{-r} \sqrt{1- \left(m-\beta^2 e^{-2r})\right)e^{-2r}}
\end{displaymath}

\begin{figure}[b]
{\includegraphics[scale=0.4,angle=-90]{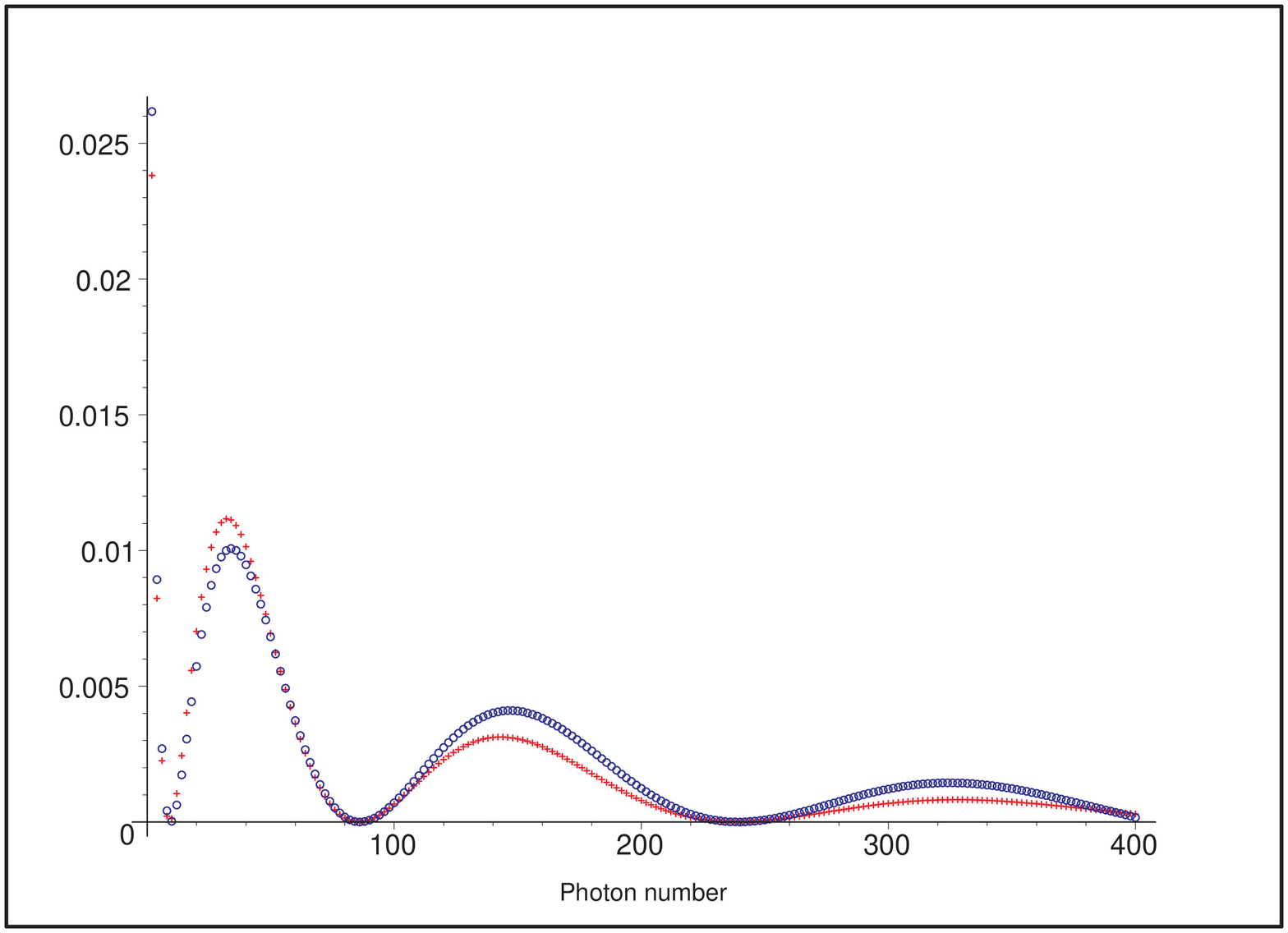} \caption{
$P_{m}$ with $m$ even for a two photon coherent state with $\beta = 5.1$ and
$r=3$.Circles: Exact result. Crosses: Approximate result.}
\label{fig11}} {\includegraphics[scale=0.4,angle=-90]{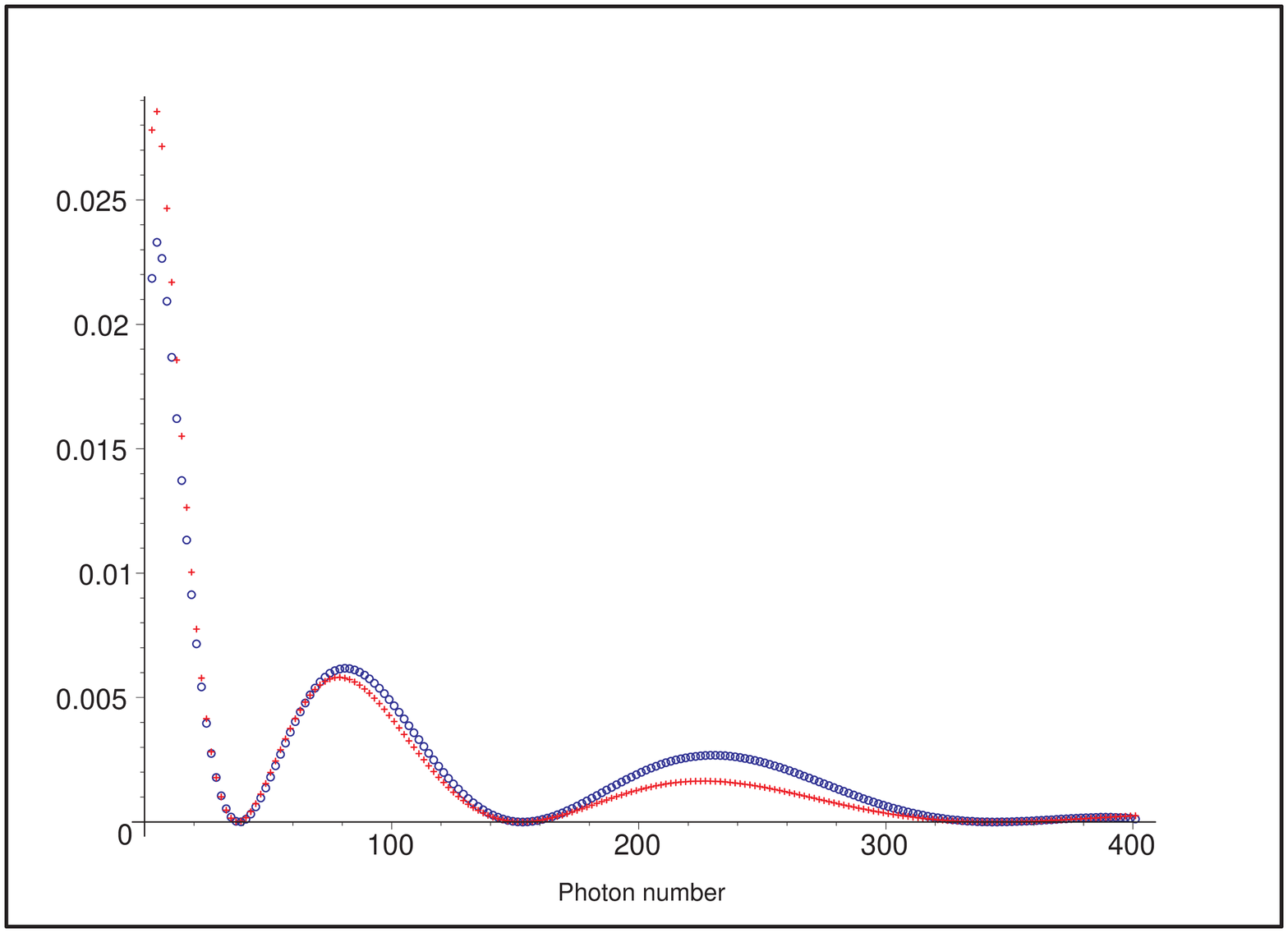}
\caption{$P_{m}$ with $m$ odd for a two photon coherent state with $\beta =
5.1$ and $r=3$.Circles: Exact result. Crosses: Approximate
result.} \label{fig12}}
\end{figure}

Figures (\ref{fig11}) and (\ref{fig12}) show the comparison
between the exact and approximate results for the photon
statistics of a two photon coherent state.

To get a geometrical interpretation   of the two branches
in the photon distribution consider again the phase evaluated at (\ref{ekis2bc}).
It may be rewritten in the form,
\begin{equation}
\psi_{\beta,r} = m \phi -X_2 Y_2 \left( \frac{e^{r}}{\cosh (r)}-
\frac{\sinh (r)}{\cosh (r)} \right) =   m \phi -X_2 Y_2
\end{equation}
From this expression follows that the phase $\psi_{\beta,r}$ is
represented by the shadowed area in Figure (\ref{fig9}). For high
values of  $r$ this phase may be written as,
\begin{equation}
\psi_{\beta,r} = m\pi/2 -2 X_2 Y_2
\end{equation}
In this case the probability is given by,
\begin{equation}
P_m = \frac{A_m}{\pi} \cos^2 (\psi_{\beta,r}) = \frac{A_m}{2 \pi} \left[1+ (-1)^m \cos ( 4 X_2 Y_2)\right]
\end{equation}
which explains the differences in the probabilities of odd and even  photon numbers.

\section{Conclusion}
In the approach to quantum phase space interference presented in
this paper we show a general setup, based in the properties of
Husimi's $Q(\alpha)$ function, for the representation of quantum
states as regions in phase space with a  prescription for
assigning phases to the different overlapping regions with a has
geometrical input. We illustrate the method discussing the photon
number distribution of the displaced number states and of the two
photon coherent states. For the latter at high squeezing the
distribution we show that the distribution develops different
oscillating behaviors for odd and even photon numbers. This effect
(also displayed by squeezed states)  is understood in terms of
phase space interference and maybe, in principle, suitable for
experimental verification.

\section{Acknowledgments}
J.S thanks the Abdus Salam International Centre for Theoretical
Physics for a Regular Associate fellowship. This work was
supported by Did-Usb grant Gid-30 and by Fonacit grant
G-2001000712.

\end{document}